\documentclass[aps,prl,reprint,superscriptaddress,showpacs]{revtex4-1}

\usepackage[pdftex]{graphicx}
\setkeys{Gin}{clip=true,draft=false}
\DeclareGraphicsExtensions{.pdf}
\usepackage[cmex10]{amsmath}
\usepackage{amsthm}
\usepackage{amsfonts}
\usepackage[cspex,bbgreekl]{mathbbol}
\begin{document}

\title{Memcomputing NP-complete problems in polynomial time using polynomial resources and collective states}


\author{Fabio L. Traversa}
\email[]{fabio.traversa@polito.it}
\affiliation{Department of Physics, University of California, San Diego, La Jolla, California 92093, USA}
\affiliation{Department of Electronics and Telecommunications, Politecnico di Torino, Turin, Italy}

\author{Chiara Ramella}
\email[]{chiara.ramella@polito.it}
\affiliation{Department of Electronics and Telecommunications, Politecnico di Torino, Turin, Italy}

\author{Fabrizio Bonani}
\email[]{fabrizio.bonani@polito.it}
\affiliation{Department of Electronics and Telecommunications, Politecnico di Torino, Turin, Italy}

\author{Massimiliano Di Ventra}
\email[]{diventra@physics.ucsd.edu}
\affiliation{Department of Physics, University of California, San Diego, La Jolla, California 92093, USA}

\date{\today}

\begin{abstract}
Memcomputing is a novel non-Turing paradigm of computation that uses interacting memory cells (memprocessors for short) to store and process information on the same physical platform\cite{13_memcomputing}. It was recently proved mathematically that memcomputing machines have the same computational power of non-deterministic Turing machines\cite{traversa_UMM}. Therefore, they can solve \emph{NP}-complete problems in polynomial time and, using the appropriate architecture, with resources that only grow polynomially with the input size. The reason for this computational power stems from properties inspired by the brain and shared by any universal memcomputing machine, in particular intrinsic parallelism and information overhead\cite{traversa_UMM}, namely the capability of compressing information in the {\it collective state} of the memprocessor network. Here, we show an experimental demonstration of an actual memcomputing architecture that solves the \emph{NP}-complete version of the subset-sum problem in only one step and is composed of a number of memprocessors that scales linearly with the size of the problem. We have fabricated this architecture using standard microelectronic technology so that it can be easily realized in any laboratory setting. Even though the particular machine presented here is eventually limited by noise--and will thus require error-correcting codes to scale to an arbitrary number of memprocessors--it represents the first proof-of-concept of a machine capable of working with the collective state of interacting memory cells, unlike the present-day single-state machines built using the von Neumann architecture.
\end{abstract}

\maketitle
\section{Introduction}
There are several classes of computational problems that require time and resources that grow exponentially with the input size when solved. This is true when these problems are solved with deterministic Turing machines, namely machines based on the well-known Turing paradigm of computation which is at the heart of any computer we use nowadays \cite{complexity_bible,computer_architecture_book}. Prototypical examples of these difficult problems are those belonging to the class that can be solved in polynomial (\emph{P}) time if a hypothetical Turing machine--named non-deterministic Turing machine--could be built. They are classified as non-deterministic polynomial (\emph{NP}) problems, and the machine is hypothetical because, unlike a deterministic Turing machine, it requires a fictitious ``oracle'' that chooses which path the machine needs to follow to get to an appropriate state \cite{36_turing,turing_book,complexity_bible}. As of today, no one knows whether \emph{NP} problems can be solved in polynomial time by a deterministic Turing machine \cite{computational_complexity_book,NP_optical}. If that were the case we could finally provide an answer to the most outstanding question in computer science, namely whether \emph{NP$=$P} or not\cite{complexity_bible}.

Very recently a new paradigm, named {\it memcomputing} \cite{13_memcomputing} has been advanced. It is based on the brain-like notion that one can process and store information within the {\it same} units (memprocessors) by means of their mutual interactions. This paradigm has its mathematical foundations on an ideal machine, alternative to the Turing one, that was formally introduced by two of us (FT and MD) and dubbed {\it universal memcomputing machine} (UMM) \cite{traversa_UMM}. Most importantly, it has been proved mathematically that UMMs have the same computational power of a non-deterministic Turing machine \cite{traversa_UMM}, but unlike the latter, UMMs are fully deterministic machines and, as such, they can actually be fabricated. A UMM owes its computational power to three main properties: {\it intrinsic parallelism}--interacting memory cells simultaneously and collectively change their states when performing computation; {\it functional polymorphism}--depending on the applied signals, the same interacting memory cells can calculate different functions; and finally {\it information overhead}--a group of interacting memory cells can store a quantity of information which is not simply proportional to the number of memory cells itself.

These properties ultimately derive from a different type of architecture: the topology of memcomputing machines is defined by a network of interacting memory cells (memprocessors), and the dynamics of this network are described by a {\it collective state} that can be used to store and process information simultaneously. This collective state is reminiscent of the collective (entangled) state of many qubits in quantum computation, where the entangled state is used to solve efficiently certain types of problems such as factorization \cite{QI_bible}. Here, we prove experimentally that
such collective states can also be implemented in classical systems by fabricating appropriate networks of memprocessors, thus creating either linear or non-linear combinations out of the states of each memprocessor. The result is the first proof-of-concept machine able to solve an \emph{NP}-complete problem in polynomial time using collective states.


The experimental realization of the memcomputing machine presented here, and theoretically proposed in Ref. \cite{traversa_UMM}, can solve the \emph{NP}-complete \cite{Karp_10} version of the subset-sum problem (SSP) in polynomial time with polynomial resources. This problem is as follows: if we consider a finite set $G\subset\mathbb{Z}$ of cardinality $n$, is there a non-empty subset $K\subseteq G$ whose sum is a given integer number $s$? As we discuss in the following paragraphs, the machine we built is analog and hence would be scalable to very large numbers of memprocessors only in the absence of noise or using some error-correcting codes. This problem derives from the fact that in the present realization we use the frequencies of the collective state to encode information and, to maintain the energy of the system bounded, the amplitudes of the frequencies are dampened exponentially with the number of memprocessors involved. However, this latter limitation is due to the particular choice of encoding the information in the collective state, and could be overcome by employing other realizations of memcomputing machines that are digital and using error-correcting codes. For example in Ref. \cite{traversa_UMM} two of us (FT and MD) proposed a different way to encode a quadratic information overhead in a network of memristors that is not subject to this energy bound.

Another example in which information overhead does not need exponential growth of energy is again quantum computing.
For instance, a close analysis of the Shor\rq{}s algorithm \cite{Shor_1} shows that the collective state of the machine implements all at once (through the superposition of quantum states) an exponential number of states, each one with the same probability that decreases exponentially with the number of qubits involved. Successively, the quantum Fourier transform reorganizes the probabilities encoded in the collective state and ``selects'' those that actually solve the implemented problem (the prime factorization in the case of the Shor\rq{}s algorithm). It is worth noticing, however, that also quantum computing algorithms necessarily require error-correcting codes for their practical implementation due to the several unavoidable sources of noise \cite{QI_bible}.

Here, it is also worth stressing that our results do not answer the \emph{NP$=$P} question, since the latter has its solution only within the Turing-machine paradigm: although a UMM is Turing-complete \cite{traversa_UMM}, it is not a Turing machine. In fact, (classical) Turing machines employ states of single memory cells and do not use collective states. Finally, we mention that other unconventional approaches to the solution of NP-complete problems have been proposed \cite{NP_optical, NP_optical2, NP_quantum, NP_molecular, NP_DNA,NP_light}, however none of them reduces the computational complexity or requires physical resources not exponentially growing with the size of the problem. On the contrary, our machine can solve an \emph{NP}-complete problem with only polynomial resources. As anticipated, this last claim is strictly valid for an arbitrary large input size only in the absence of noise.

\section{Results}
\subsection{Implementing the SSP.}


\begin{figure*}
\begin{center}
\includegraphics[width=1.7\columnwidth]{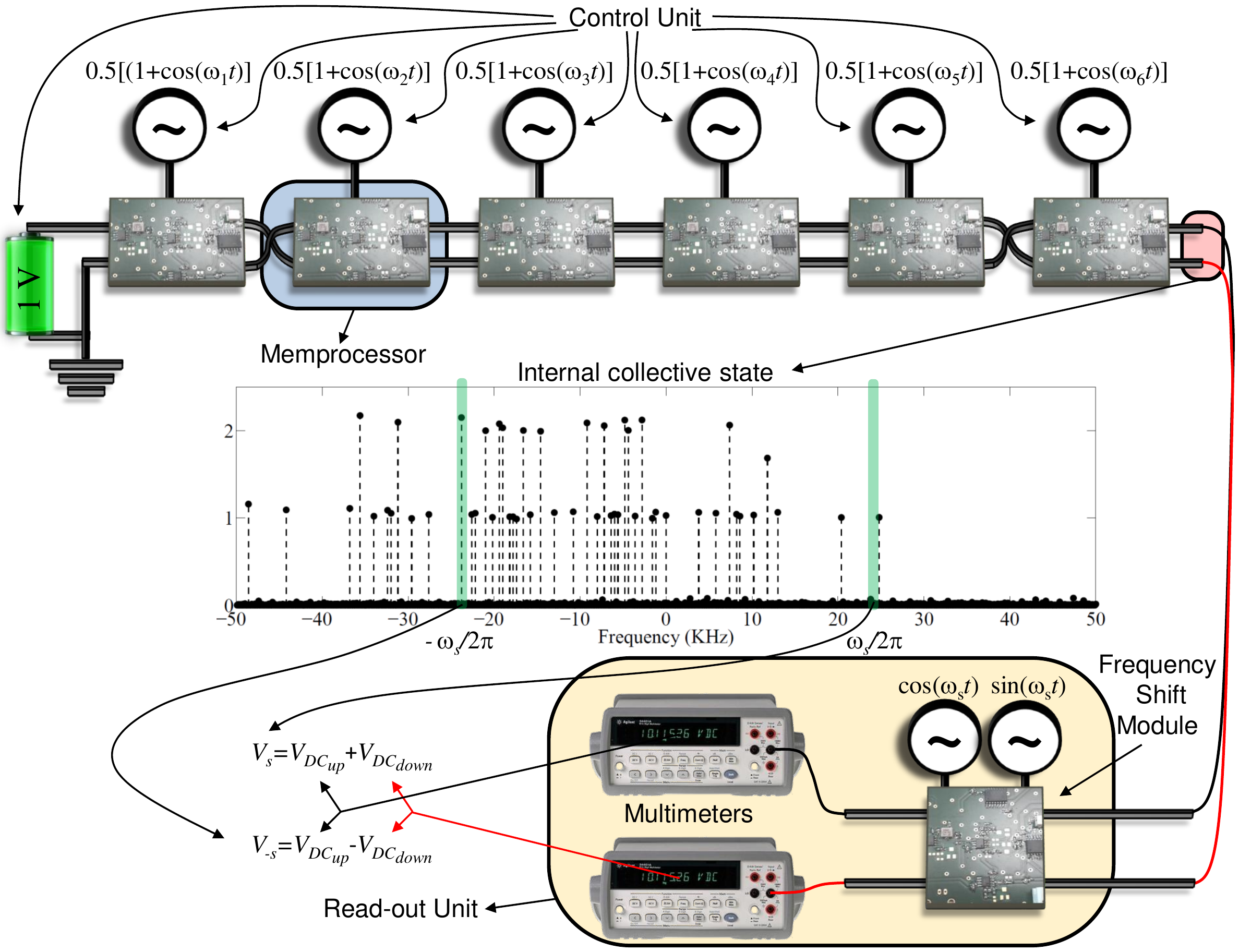}
\end{center}
\caption{\label{scheme}Scheme of the memcomputing architecture used in this work to solve the subset-sum problem. The central spectrum has been obtained by the discrete Fourier transform of the experimental output of a network of 6 memprocessors encoding the set $G=\{130,$ $-130,$ $-146,$ $-166,$ $-44,$ $118\}$ with fundamental frequency $f_0=100$~\textrm{Hz}.}%
\end{figure*}

\begin{figure}
\begin{center}
\includegraphics[width=\columnwidth]{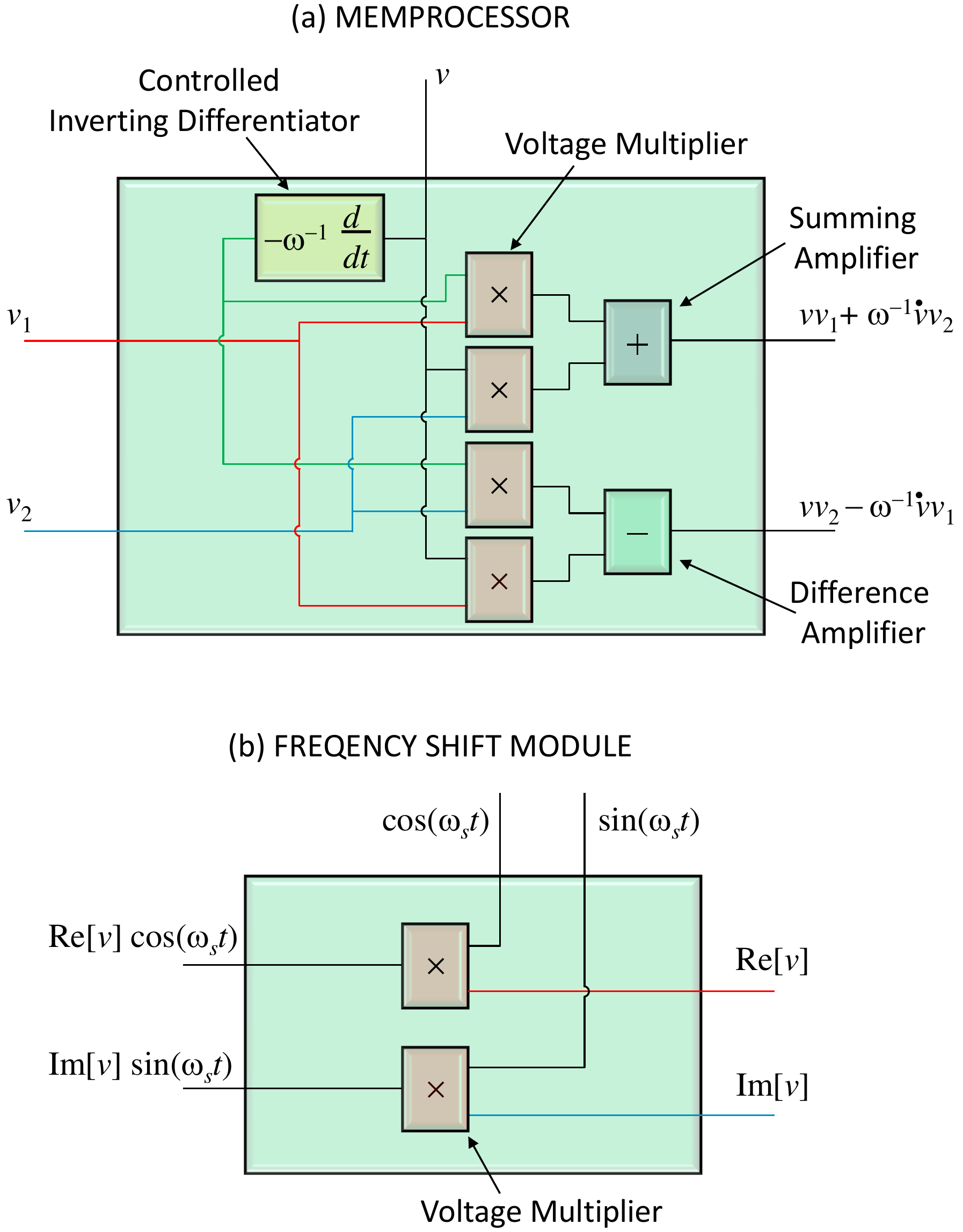}
\end{center}
\caption{\label{memprocessor}(a) Simplified schematic of the memprocessor architecture used in this work to solve the subset-sum problem (more details can be found in the Supplementary Information). (b) Schematic of the frequency shift module.}%
\end{figure}


The machine we built to solve the SSP is a particular realization of a UMM based on the \emph{memcomputing architecture} described in Ref. \cite{traversa_UMM}, namely it is composed of a control unit, a network of memprocessors (computational memory) and a read-out unit as schematically depicted in Figure~\ref{scheme}. The control unit is composed of generators applied to each memprocessor. The memprocessor itself is an electronic module fabricated from standard electronic devices, as sketched in Figure~\ref{memprocessor} and detailed in the Materials and Methods section. Finally, the read-out unit is composed of a frequency shift module and two multimeters. All the components we have used employ commercial electronic devices.


The control unit feeds the memprocessor network with sinusoidal signals (that represent the input signal of the network) as in Figure~\ref{scheme}. It is simple to show that the collective state of the memprocessor network of this machine (that can be read at the final terminals of the network) is given by the real (up terminal) and imaginary (down terminal) part of the function
\begin{equation}
g(t)=2^{-n}\prod_{j=1}^{n}(1+\exp[i\omega_jt])\label{collective_state}
\end{equation}
where $n$ is the number of memprocessors in the network and $i$ the imaginary unit (see Materials and Methods or Ref. \cite{traversa_UMM}). If we indicate with $a_j\in G$ the $j$-th element (integer with sign) of $G$, and we set the frequencies as $\omega_j=2\pi a_j f_0$ with $f_0$ the fundamental frequency equal for any memprocessor, we are actually encoding the elements of $G$ into the memprocessors through the control-unit feeding frequencies. Therefore, the frequency spectrum of the collective state \eqref{collective_state} (or more precisely the spectrum of $g(t)-2^{-n}$) will have the harmonic amplitude, associated with the normalized frequency $f=\omega/(2\pi f_0)$, proportional to the number of subsets $K\subseteq G$ whose sum $s$ is equal to $f$. In other words, if we read the spectrum of the collective state \eqref{collective_state}, the harmonic amplitudes are the solution of the subset sum problem for any $s$. From this first analysis we can make the following considerations.

\textit{Information overhead:} the memprocessor network is fed by $n$ frequencies encoding the $n$ elements of $G$, but the collective state \eqref{collective_state} encodes {\it all} possible sums of subsets of $G$ into its spectrum. It is well known \cite{computational_complexity_book} that the number of possible sums $s$ (or equivalently the scaled frequencies $f$ of the spectrum) can be estimated in the worst case as $O(A)$ where $A=\max[\sum_{a_j>0}a_j,-\sum_{a_j<0}a_j]$. Obviously $A$ (sometimes called the \emph{capacity of the problem}) has exponential growth \cite{algorithms_book} on the minimum number $p$ of bits used to represent the elements of $G$ ($p$ is called precision of the problem and we have $A=O(2^p)$, if we take the precision in bits). Thus the spectrum of the collective state \eqref{collective_state} encodes an information overhead that grows {\it exponentially} with the precision of the  problem.

\textit{Computation time:} the collective state \eqref{collective_state} is a periodic function of $t$ with minimum period $T=1/f_0$ because all frequencies involved in \eqref{collective_state} are multiples of the fundamental frequency $f_0$. Therefore, $T$ is the minimum time required for computing the solution of the SSP within the memprocessor network and so it can be interpreted as the computation time of the machine. However, this computation time is {\it independent} of both $n$ and $p$.

\textit{Energy expenditure:} the energy required to compute the SSP can be estimated as that quantity proportional to the energy of the collective state in one period $E=\int_0^T|g(t)|^2dt$. By using \eqref{collective_state} we have $E\leq\int_0^Tdt\leq1/f_0$, so also the energy needed for the computation is independent of both $n$ and $p$. It is  worth remarking here that, in order to keep the energy bounded, all generators have the coefficient 0.5 (see Figure~\ref{scheme}) then introducing (see Materials and Methods) the factor $2^{-n}$ in Eq.~\eqref{collective_state}. This means that all frequencies involved in the collective state \eqref{collective_state} are dampened by the factor $2^{-n}$. In the case of the ideal machine, i.e., a noiseless machine, this would not represent an issue because no information is lost. On the contrary, when noise is accounted for, the exponential factor represents the hardest limitation of the experimentally fabricated machine, which we reiterate is a
technological limit for this particular realization of a memcomputing machine but not for all of them.

\subsection{Reading the SSP solution}

With this analysis we have proven that the UMM represented in Figure~\ref{scheme} can solve the SSP with $n$ memprocessors, a control unit formed by $n+1$ generators and taking a time $T$ and an energy $E$ independent of both $n$ and $p$. Therefore, at first glance, it seems that this machine (without the read-out unit) can solve the SSP using only resources polynomial (specifically, linear) in $n$. However, we need one more step: we have to {\it read} the result of the computation. Unfortunately, we cannot simply read the collective state \eqref{collective_state} using, e.g., an oscilloscope and performing the Fourier conversion. This is because the most optimized algorithm to do this (see Materials and Methods and Ref. \cite{traversa_UMM}) is exponential in $p$, i.e., it has the same complexity of standard dynamic programming \cite{algorithms_book}.

However, a solution to this problem can be found by just using standard electronics to implement a read-out unit capable of extracting the desired frequency amplitude without adding any computational burden. In Figure~\ref{scheme} we sketch the  read-out unit we have used. It is composed of a frequency-shift module and two multimeters. The frequency shift module is in turn composed of two voltage multipliers and two sinusoidal generators as depicted in Figure~\ref{memprocessor} and it works as follows. If we connect to one multiplier the real part of a complex signal $v(t)$ and to the other multiplier the imaginary part, we obtain at the outputs $v_{up}(t)=\mathrm{Re}[v(t)](\exp[i\omega_s t]+\exp[-i\omega_s t])/2$ and $v_{down}(t)=-i\mathrm{Im}[v(t)](\exp[i\omega_s t]-\exp[-i\omega_s t])/2$. The sum and difference of the two outputs are $v_{up}+v_{down}=\mathrm{Re}[v(t)\exp[-i\omega_s t]]$ and $v_{up}-v_{down}=\mathrm{Re}[v(t)\exp[i\omega_s t]]$. From basic Fourier calculus, $v(t)\exp[-i\omega_s t]$ and $v(t)\exp[i\omega_s t]$ are the frequency spectrum shifts of $\omega_s/(2\pi)$ and $-\omega_s/(2\pi)$ for the function $v(t)$, respectively. Consequently, if we read the DC voltages $V_{DC_{up}}$ and $V_{DC_{down}}$ of $v_{up}(t)$ and $v_{down}(t)$ using two multimeters and perform the sum $V_{DC_{up}}+V_{DC_{down}}$ and difference $V_{DC_{up}}-V_{DC_{down}}$ we obtain the amplitudes $V_s$ and $V_{-s}$ of the harmonics at the frequencies $\omega_s/(2\pi)$ and $-\omega_s/(2\pi)$, respectively (see Figure~\ref{scheme}).



Hence, by feeding the frequency shift module with the $\mathrm{Re}[g(t)]$ and $\mathrm{Im}[g(t)]$ from \eqref{collective_state}, reading the output with the two multimeters, and performing sum and difference of the final outputs we obtain the harmonic amplitude for a particular normalized frequency $f$ according to the external frequency $\omega_s$ of the frequency-shift module. In other words, without adding any additional computational burden and time, we can solve the SSP for a given $s$ by properly setting the external frequency of the frequency-shift module.

It is worth noticing that, if we wanted to {\it simulate} our machine by including the read-out unit the computational complexity would be $O(2^p)$ (close to the standard dynamic programming that is $O(n2^p)$ \cite{algorithms_book}). In fact, from the Nyquist-Shannon sampling theorem\cite{wiley_enc}, the minimum number of samples of the shifted collective state (i.e,. the outputs of the frequency-shift module) must be equal to the number of frequencies of the signal (in our case of $O(2^p)$) in order to accurately evaluate even one of the harmonic amplitudes \cite{Goertzel}. The last claim can be intuitively seen from this consideration: the DC voltage $V_{DC_{up}}$ must be calculated in the simulation by evaluating the integral $V_{DC_{up}}=T^{-1}\int_0^Tv_{up}(t)dt$ and this requires at least $O(2^p)$ samples for an accurate evaluation \cite{wiley_enc}. On the other hand, the multimeter of the hardware implementation, being essentially a narrow low-pass filter, performs an analog implementation of the integral over a continuous time interval $T$ (independent of $n$ and $p$), directly providing the result, thus avoiding the need of sampling the waveform and computing the integral.

\begin{figure*}
\begin{center}
\includegraphics[width=1.7\columnwidth]{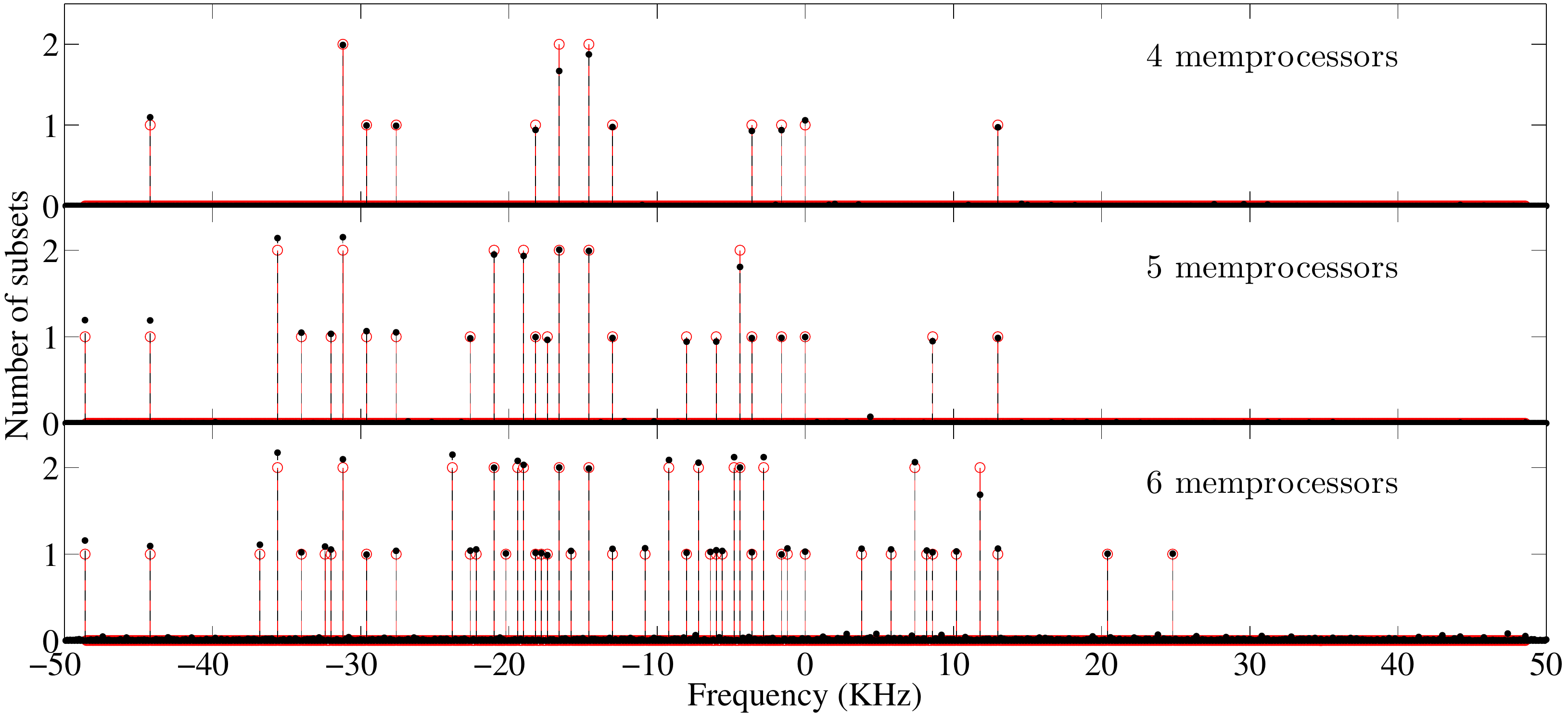}
\end{center}
\caption{\label{spectrum}Spectra of internal collective state of three different networks with fundamental frequency $f_0=100$~\textrm{Hz}. 4--memprocessor network encodes the set $G=\{130,$ $-130,$ $-146,$ $-166\}$. 5--memprocessor network encodes $G=\{130,$ $-130,$ $-146,$ $-166,$ $-44\}$. 6--memprocessor network encodes $G=\{130,$ $-130,$ $-146,$ $-166,$ $-44,$ $118\}$.}%
\end{figure*}

\begin{table}
\centering
\begin{tabular}{|c|c|c|c|c|c|c|c|}
\hline
{$|s|$} &{$\omega_s/(2\pi)$} & {$V_{DC_{up}}$} & {$V_{DC_{down}}$}&{$V_s$} &{$V_{-s}$}& {\#subs. } & {\#subs.} \\
{} &{\small{[KHz]}} & {\small{[mV]}} & {\small{[mV]}} &{\small{[V]}} &{\small{[V]}} & {sum $s$} & {sum $-s$} \\ \hline
0 & 0 &  31.7 & 0 &  1.02 &  1.02 &  1 & 1\\ \hline
  74 &    7.4 &  15.3 & 15.0 & 1.94  & 0.02 &  2 & 0\\ \hline
130 &  13.0 &   -0.2 & 14.9 & 0.94  &-0.97 &  1 & 1\\ \hline
146 &  14.6 &  14.8 & 15.8 & -0.06 & 1.96 &  0 & 2\\ \hline
248 &  24.8 &    7.6 &   7.2 & 0.95  & 0.02 &  1 & 0\\ \hline
485 &  48.5 &   -0.4 & -0.7 & -0.07 & 0.02 & 0 & 0\\ \hline
486 &  48.6 &   -8.9 &  6.6 & -0.14 & 0.99 &  0 & 1\\ \hline
\end{tabular}
\caption{\label{table}Measurements from the read-out unit of Figure~\ref{scheme} for a 6--memprocessor network with fundamental frequency $f_0=100$~\textrm{Hz} encoding the set $G=\{130,$ $-130,$ $-146,$ $-166,$ $-44,$ $118\}$. In the fifth and sixth columns the voltages are respectively given by $V_s=2^6(V_{DC_{up}}+V_{DC_{down}})$ and $V_{-s}=2^6(V_{DC_{up}}-V_{DC_{down}})$. The last two columns are the analytical results.}
\end{table}

In Figure~\ref{spectrum} the absolute value of the spectrum of the collective state for networks of 4, 5 and 6 memprocessors is compared with the theoretical results given by the spectrum of \eqref{collective_state} (see Materials and Methods for more details on the hardware and measurement process). Non-idealities of the circuit and electronic noise in general are the sources of the small discrepancies with respect to the theoretical results.
Nevertheless, the machine we fabricated demonstrates that using the collective state of all memprocessors, instead of the uncoupled states of the individual memory units, we can carry out difficult computing tasks (NP-complete problems) with polynomial resources. Finally, in Table~\ref{table}, the measurements at the read-out circuit are listed for different harmonics for a 6-memprocessor network. The precision is up the third digit as can be seen from the comparison with the theoretical results.

\section{Discussion}
\subsection{Scalability and Error Correction}

As anticipated, the machine we have built would ultimately be limited by unavoidable noise sources, thus requiring error-correcting codes. However we prove here that under the assumption of low noise, additive white noise does not affect the machine output. Therefore, only non-idealities of the devices and colored noise represent a limit for this particular machine. To analyze this issue we consider first the simplest case of independent white Gaussian additive noise sources for each memprocessor. Under this hypothesis each memprocessor has complex noisy input of the form $v_{j_{in}}=2^{-1}(1+\exp[i\omega_j t])+\varepsilon_j(t)$. When we connect two memprocessors, the output at the terminals of the second memprocessor is given by
\begin{align}
v_{2_{out}} =&v_{1_{in}}v_{2_{in}}=  2^{-2}(1+e^{i\omega_{1}t})(1+e^{i\omega_{2}t})+\nonumber\\
 +& 2^{-1}(1+e^{i\omega_{1}t})\varepsilon_{2}(t)+2^{-1}(1+e^{i\omega_{2}t})\varepsilon_{1}(t)+\nonumber\\
 +&\varepsilon_{1}(t)\varepsilon_{2}(t).
\end{align}

If the noise term is small enough (low noise components) the quantity $\varepsilon_{1}(t)\varepsilon_{2}(t)$ is negligible compared to the other terms and can be neglected. Following the same steps, the network composed of $n$ memprocessors has collective state (under low noise condition) of the form
\begin{align}
g(t)=&v_{n_{out}}=2^{-n}\prod_{j=1}^{n}(1+e^{i\omega_{j}t})+\nonumber\\
+&2^{-n+1}\sum_{k=1}^{n}\varepsilon_{k}(t)\prod_{\substack{j=1\\j\neq k}}^{n}(1+e^{i\omega_{j}t})=g_{S}(t)+g_{N}(t) \label{noisy_collective_state}
\end{align}

Using the expression (\ref{noisy_collective_state}) we can calculate the signal-to-noise-ratio as the ratio between the power of the signal and the
power of the noise. The signal power (neglecting noise) is simply given by $S=E=\int_{0}^{T}|g_{S}(t)|^{2}dt\approx T$ and the noise power is given by $N=\int_{0}^{T}E[|g_{N}(t)|^{2}]dt$, where $E[\cdot]$ is the expectation value operator. Since the noise sources are independent we have $E[|g_{N}(t)|^{2}]\approx\sum_{k=1}^{n}E[|\varepsilon_{k}(t)|^{2}]$, and being the sources white $E[|g_{N}(t)|^{2}]$ is independent of time.
Finally, assuming $E[|\varepsilon_{k}(t)|^{2}]=E[|\varepsilon_{j}(t)|^{2}]$ for any $j$ and $k$ we have $E[|g_{N}(t)|^{2}]\approx TnE[|\varepsilon|^{2}]$ and the signal-to-noise ratio reads
\begin{equation}
\frac{S}{N}\approx\frac{1}{nE[|\varepsilon|^{2}]}\label{SN}\,.
\end{equation}
However, being the noise withe, the noise power spectrum, according to \eqref{noisy_collective_state}, is distributed among the different harmonics with weights given by the coefficients in \eqref{noisy_collective_state} that are exponentially decreasing with $n$. Therefore, the signal to noise ratio for each harmonic has the same order of magnitude of \eqref{SN} and hence the withe noise (under conditions of low noise) does not affect the machine when we scale it up.

On the other hand, non-idealities, colored and $1 \backslash f$ noise and other non-Gaussian noise sources, have spectra that are not distributed on the harmonics as withe noise. For example, the $1\backslash f$ noise accumulates on the output harmonic during the process of measurement. Nevertheless, if the total noise can be considered low, the equation \eqref{noisy_collective_state} is still valid but the ratio \eqref{SN} should be taken frequency dependent. In this case we can apply the Shannon\rq{}s noisy-channel coding theorem \cite{Shannon} with our machine interpreted as a noisy channel with capacity $C$. The input is given by the control unit and the output is the collective state $g(t)$. The Shannon theorem then states that for any $\varepsilon>0$ and $R<C$, there exists a code of length $N$ and rate $\geq R$ and a decoding algorithm, such that the maximal probability of block error is $\leq\varepsilon$.

In addition, using the Shannon--Hartley theorem \cite{Communication_book} we have that for frequency dependent noise the capacity $C$ can be calculated as
\begin{equation}
C=\int^B_0\log_{2}\left(  1+\frac{S(f)}{N(f)}\right)df,
\end{equation}
where $B$ is the bandwidth of the channel. In our case, the lower bound of $B$ can be taken as linear in the number of memprocessors, i.e., $B=B_0n$ with $B_0$ a constant, and $S/N$ is given by (\ref{SN}). We finally have (for large $n$)
\begin{equation}
C\approx \int^B_0\log_{2}\left(  1+\frac{1}{nE[\varepsilon^{2}(f)]}\right)df  \approx \frac{B_0}{\overline{E[\varepsilon^{2}]}\ln2}.
\end{equation}
where $\overline{E[\varepsilon^{2}]}^{-1}=\lim_{n\rightarrow \infty} \int^{nB_0}_0 E[\varepsilon^{2}(f)]^{-1}df$.

We therefore conclude that our machine compresses data in an exponential way with constant capacity. At the output we have the exponential decreasing probability of finding one solution of the subset sum problem when we implement brute force the algorithm, i.e., without any error-correcting coding. However, from Shannon theorem, there exists a code that allows us to send the required information with bounded error. The question is then whether there is a polynomial code that accomplishes this task. We briefly discuss here the question heuristically. If our machine were Turing-like, the polynomial code could exist only if $NP=P$. Instead, our machine is not Turing-like, and the channel can perform an exponential number of operations at the same time. Since this is similar to what quantum computing does when solving difficult problems such as factorization,  and we know that for quantum computing polynomial correcting codes do exist \cite{QI_bible}, we expect similar coding can be applied to our specific machine as well.

\section{Conclusions}
In summary, we have demonstrated experimentally a deterministic memcomputing machine that is able to solve an $NP$-complete problem in polynomial time (actually in one step) using only polynomial resources. The actual machine we built clearly suffers from technological limitations due to unavoidable noise that impair its scalability. These limitations derive from the fact that we encode the information directly into frequencies, and so ultimately into energy. This issue could, however, be overcome either using error correcting codes or with other UMMs that use other ways to encode such information and are digital at least in their input and output. Irrespective, this machine represents the first experimental realization of a UMM that uses the collective state of the whole memprocessor network to exploit the information overhead theoretically introduced in Ref.~\cite{traversa_UMM}.
Finally, it is worth mentioning that the machine we have fabricated is not a general purpose one. However, other realizations of UMMs are general purpose and can be easily built with available technology\cite{DCRAM,07_waser,08_strukov,09_memory_materials,12_grollier}. Their practical realization would thus be a powerful alternative to current Turing-like machines.

\section{Materials and Methods}

\subsection{Experimental Design}

The operating frequency range of the experimental setup needs to be discussed with the measurement target in mind. For the measurement of the entire collective state, the limiting frequency is due to the oscilloscope we use to sample the full signal at the output of the memprocessor network. On the other hand, the measurement of some isolated harmonic amplitudes using the read-out unit, transfers this bottleneck to the voltage generators and internal memprocessor components. It is worth stressing that measuring the collective state is {\it not} the actual target of our work because it has the same complexity of the standard algorithms for the SSP as discussed in the Results section and in Ref. \cite{traversa_UMM}. Here, for completeness, we provide measurements of the collective state only to prove that the setup works properly. The actual frequency range of the setup is discussed in the next section.

\subsection{Setup frequency range}   

Let us consider $a_j\in G$, and the integer
\begin{equation}
A=\max \left\{-\sum_{a_j<0}a_j,\sum_{a_j>0}a_j\right\}. \label{max_freq}
\end{equation}
We also consider $f_0\in \mathbb{R}$ and we encode the $a_j$ in the frequencies by setting the generators at frequencies $f_j=|a_j|f_0$, so the maximum frequency of the collective state is $f_{\max}=Af_0\label{fmax}$. From these considerations, we can first determine the range of the voltage generators: it must allow for the minimum frequency (\emph{resolution})  
\begin{equation}
f_{g_{\min}}\leq f_0\label{fmin}
\end{equation}
and maximum frequency (\emph{bandwidth}) 
\begin{equation}
f_{g_{\max}}\geq f_0\max_G\{|a_j|\}.\label{fmax1}
\end{equation}
We employed the Agilent 33220A Waveform Generator\footnote{Note that when we use digital voltage generators we need to control the frequency with precision proportional to $p$. Therefore, generating waveforms with digital generators should require exponential resources in $p$ because of the time sampling. However, this can be in principle overcome using synchronized analog generators designed for this specific task.}, which has 1 $\mu$Hz resolution and 20 MHz bandwidth. This means that, in principle, we can accurately encode $G$ when composed of integers with a precision up to 13 digits (which is quite the same precision of the standard double-precision integers) provided that a stable and accurate external clock reference, such as a rubidium frequency standard is used. This is because the internal reference of such generators introduces a relative uncertainty on the synthesized frequency in the range of some parts-per-billion ($10^{-9}$) thus limiting the resolution at high frequency, down to few mHz at the maximum frequency. However, as anticipated, this issue can be resolved by employing an external reference providing higher accuracies. On the other hand, note that the frequency range can be, in principle, increased by using wider bandwidth generators, up to the GHz range.

Another frequency limitation concerning the maximum operating frequency is given by the electronic components of the memprocessors. In fact, the active elements necessary to implement in hardware the memprocessor modules have specific operating frequencies that cannot be exceeded. Discrete operational amplifiers (OP-AMPs) are the best candidates for this implementation due to their flexibility in realizing different types of operations (amplification, sum, difference, multiplication, derivative, etc.). Their maximum operating frequency is related to the gain-bandwidth product (GBWP). We used standard high frequency OP-AMP that can reach GBWP up to few GHz. However, such amplifiers usually show high sensitivity to parasitic capacitances and stability issues (e.g., a limited stable gain range). Typical maximum bandwidth of such OP-AMPs that ensures unity gain stability and acceptable insensitivity to parasitics are of the order of few tens of MHz, thus compatible with the bandwidth of the Agilent 33220A Waveform Generator. Therefore, we can set quantitatively the last frequency limit related to the hardware as
\begin{equation}
f_{OPAMP_{\max}}>Af_0, \label{fmax2}
\end{equation}
and ensure optimal OP-AMP functionality. Finally using \eqref{fmin}-\eqref{fmax2} we can find a reasonable $f_0$ satisfying the frequency constraints.

\subsection{Memprocessor}

The memprocessor, synthetically discussed in the Results section and sketched in Figure~\ref{memprocessor}, is shown in the pictures of Figure~S1-(a) and a more detailed circuit schematics is given in Figure~S1-(b). Each module has been realized as a single Printed Circuit Board (PCB) and connections are performed through coaxial cables with BNC terminations. According to Figure~\ref{memprocessor}, each memprocessor must perform one derivative $-\omega^{-1}_j\frac{d}{dt}$, four multiplications, one sum and one difference.
Since $v(t)=0.5[1+\cos(2\pi f_0a_jt)]$ and $\omega_j=2\pi f_0a_j$, then $-\omega^{-1}_j\dot{v}(t)=0.5\sin(2\pi f_0a_jt)$, i.e. the quadrature signal with respect to the input $v(t)$. This can be easily obtained with the simple OP-AMP-based inverting differentiator depicted in Figure~S1-(b), designed to have unitary gain at frequency $\omega_j$. Similarly, an inverting summing amplifier and a difference amplifier can be realized as sketched in Figure~S1-(b) to perform sum and difference of voltage signals, respectively. The OP-AMP selected is the Texas Instruments LM7171 Voltage Feedback Amplifier.

Implementing multiplication is slighly more challenging. OP-AMP based analog multipliers are very sensitive circuits. Therefore,  they need to be carefully calibrated. This makes a discrete OP-AMP based realization quite challenging and the integration expensive. We thus adopted a pre-assembled analog multiplier: the Texas Instrument (Burr-Brown) MPY634 Analog Multiplier, which ensures four-quadrant operation, good accuracy and wide enough bandwidth (10MHz). The only drawback of this multiplier is that the maximum precision is achieved with a gain of 0.1. Therefore, being the input signals in general quite small, this further lowering of the precision can make the output signal comparable to the offset voltages of the subsequent OP-AMP stages. For this reason, we have included in the PCB a gain stage (inverting amplifier) before each output to compensate for the previous signal inversion and lowering. These stages permit also manual offset adjustment by means of a tunable network added to the non-inverting input, as shown in the schematic. Finally a low-pass filter with corner frequency $f_c\gg Af_0$ has been added to the outputs to limit noise. Figure~S1-(a) shows a picture of one of the modules, which have been realized on a 100 mm $\times$ 80 mm PCB. The power consumption of each module is quite high since all of the 10 active components work with $\pm15$ V supply. OP-AMPs have a quiescent current of 6.5 mA, while the multipliers a quiescent current of 4 mA, yielding a total DC current of around 50 mA per module.

Finally, we briefly discuss how connected memprocessors work. From Figure~\ref{memprocessor}, if we have $v(t)=0.5[1+\cos(\omega t)]$, $v_1=\mathrm{Re}[f(t)]$ and $v_2=\mathrm{Im}[f(t)]$ for an arbitrary complex function $f(t)$, at the output of the memprocessor we will find 
\begin{align}
v_{1_{out}}=0.5[1+\cos(\omega t)]&\mathrm{Re}[f(t)]-0.5\sin(\omega t)\mathrm{Im}[f(t)]=\nonumber\\
&=\mathrm{Re}[0.5\left(1+e^{i\omega t}\right)f(t)]\\
v_{2_{out}}=0.5[1+\cos(\omega t)]&\mathrm{Im}[f(t)]+0.5\sin(\omega t)\mathrm{Re}[f(t)]=\nonumber\\
&=\mathrm{Im}[0.5\left(1+e^{i\omega t}\right)f(t)]
\end{align}
Since we can only set positive frequencies for the generators, in order to encode negative frequencies (i.e., a negative $a_j$) we can simply invert the input and output terminals as depicted in Figure~S1-(c). Therefore, if we set $f(t)=1$ for the first memprocessor, i.e., $v_1=1$ and $v_2=0$, at the output of the first memprocessor we will find the real and imaginary parts of $0.5(1+\exp[i\omega_1t])$ that will be the new $f(t)$ for the second memprocessor. Proceeding in this way we find the collective state \eqref{collective_state} at the end of the last memprocessor.

\subsection{Analysis of the Collective State}

In order to test if the memprocessor network correctly works, we have carried out the measurement of the full collective state $g(t)$ at the end of the last memprocessor. This task requires an extra discussion on the operating frequency range. Indeed, the instrument we used to acquire the output waveform is the LeCroy WaveRunner 6030 Oscilloscope. The measurement process consists in acquiring the output waveforms and apply the FFT in software. Being the collective state a purely periodic signal, i.e., a signal containing only frequencies multiples of $f_0$ and with known maximum frequency $f_{\max}=Af_0$, from the discrete Fourier transform theory and Nyquist-Shannon sampling theorem\cite{traversa_UMM,wiley_enc}, we need to sample the interval $[0,1/f_0]$ into $N=2f_{\max}/f_0+1$ subintervals of width $\Delta t=(Nf_0)^{-1}$ to compute the \emph{exact} spectrum of $g(t)$. In other words we need samples $g(t_j)$ with $t_j=k\Delta t$ and $k=0,...,N-1$ to compute the exact spectrum of $g(t)$. Therefore, with the oscilloscope we must be able to acquire at least $N+1$ samples of $g(t)$ into the time interval $[0,1/f_0]$.

This relationship turns out to be a constraint on the usable frequency range since we must perform the measurement in a reasonable time, and we cannot exceed the maximum sampling frequency of the instrument that we use for acquisition, nor its maximum memory capability. 
In our experimental proof, the LeCroy WaveRunner 6030 Oscilloscope is characterized by 350~MHz bandwidth, 2.5~GSa/s sampling rate and $10^5$ discretization points when saving waveforms in ascii format (i.e., $N_{O_{\max}}=10^5$). The bandwidth of the oscilloscope is very large thus it is not really a constraint, while the limit $N_{O_{\max}}$ is, in fact we have the constraint
\begin{equation}
f_{max}\leq\frac{f_0}{2}(N_{O_{\max}}-1)\,.
\end{equation}
With this value, choosing $f_0=1$~Hz allows to have $A\leq(10^5-1)/2$, $f_{\max}\lesssim50$~KHz and requires 1 s measurement time which is a quite long time in electronics. Therefore, without varying $A$, we can choose larger $f_0$ that allows for a smaller measurement time. We choose $f_0=100$~Hz which means a measurement time of only few tens of ms, and $f_{\max}\lesssim5$~MHz.

\subsection{Experimental set-up}

The laboratory set-up we have employed is sketched in Fig.~S2-(a), while Fig.~S3 reports a picture of the same. The order of cascade connection of the memprocessors is arbitrary. For this test, we ordered the module such that $G=\{130,-130,-146,-166,-44,118\}$ (see Figure~S2-(a)) in order to have the two memprocessors related to the two positive numbers (130 and 118) at the beginning and at the end of the chain, respectively, thus minimizing the number of the \lq\lq{}swapped\rq\rq{} connections (see Figure~S1-(c)). A two-output power supply (model Agilent E3631A) is used to generate both the $0$ and $1$~V at the inputs of the first memprocessor and the $\pm15$ V supply for all the modules (parallel connection). The input $v(t)$ of each module is generated by a Agilent 33220A waveform generator, while the output is observed through both an oscilloscope (model LeCroy Waverunner 6030, see Figure~S2-(c)) and a multimeter (model Agilent 34401A). In particular, the oscilloscope is used for the AC waveform, while the multimeter mesured the DC component in order to avoid errors due to the oscilloscope probes which are very inaccurate at DC and may show DC offsets up to tens of mV.

Another issue concerns the synchronization of the generators. The six generators we use must share the same time base and they must have the same starting instant (the $t=0$ instant) when all the cosine waveforms must have amplitude 0.5 V. To have a common time base we used the 10 MHz time base signal of one of the generators (master) and we connected the master output signal to all the other (slave) generators at the 10 MHz input. In this way they ignore their own internal time base and lock to the external one. In order to have a common $t=0$ instant we must run the generators in the \emph{infinite burst mode}. In this mode the generators produce no output signal until a trigger input is given, and then they run indefinitely until they are manually stopped. The trigger input can be external or manual (soft key): the master device is set up to expect a manual input while the slave devices are controlled by an external trigger coming from the master. Finally, in order to correctly visualize the output waveforms we must ``synchronize'' also the oscilloscope. The trigger signal of the oscilloscope must have the same frequency of the signal to be plotted, or at least one of its subharmonics. Otherwise, we see the waveform moving on the display and, in case two or more signals are acquired, we lose the information concerning their phase relation. To solve this problem we connect the external trigger input of the oscilloscope to a dedicated signal generator, producing a square waveform at frequency $f_0$, which is the greatest common divisor of all the possible frequency components of the output signals. This dedicated generator is also used as the master for synchronization. Fig.~S2-(b) shows how the generators must be connected in order to obtain the required synchronization.

\paragraph*{Acknowledgements}
The hardware realization of the machine presented here was supported by Politecnico di Torino through the Neural Engineering and Computation Lab initiative. MD acknowledges partial support from CMRR.

\bibliographystyle{unsrt}
 \bibliography{UMM_arxiv}

\begin{thebibliography}{10}

\bibitem{13_memcomputing}
Massimiliano Di~Ventra and Yuriy~V. Pershin.
\newblock The parallel approach.
\newblock {\em Nature Physics}, 9:200--202, 2013.

\bibitem{traversa_UMM}
F.~L. Traversa and M.~Di~Ventra.
\newblock Universal $\mathrm{M}$emcomputing $\mathrm{M}$achines.
\newblock {\em (preprint on arXiv:1405.0931) IEEE Transaction on Neural
  Networks and Learning Systems, DOI: 10.1109/TNNLS.2015.2391182}, 2015.

\bibitem{complexity_bible}
Michael~R. Garey and David~S. Johnson.
\newblock {\em Computers and Intractability; A Guide to the Theory of
  NP-Completeness}.
\newblock W. H. Freeman \& Co., New York, NY, USA, 1990.

\bibitem{computer_architecture_book}
John~L. Hennessy and David~A. Patterson.
\newblock {\em Computer Architecture, Fourth Edition: A Quantitative Approach}.
\newblock Morgan Kaufmann Publishers Inc., San Francisco, CA, USA, 2006.

\bibitem{36_turing}
Alan~M. Turing.
\newblock On computational numbers, with an application to the
  entscheidungsproblem.
\newblock {\em Proc. of the London Math. Soc.}, 42:230--265, 1936.

\bibitem{turing_book}
Alan~M. Turing.
\newblock {\em The Essential Turing: Seminal Writings in Computing, Logic,
  Philosophy, Artificial Intelligence, and Artificial Life, Plus The Secrets of
  Enigma}.
\newblock Oxford University Press, 2004.

\bibitem{computational_complexity_book}
Sanjeev Arora and Boaz Barak.
\newblock {\em Computational Complexity: A Modern Approach}.
\newblock Cambridge University Press, 2009.

\bibitem{NP_optical}
Kan Wu, Javier~Garc{\'\i}a de~Abajo, Cesare Soci, Perry~Ping Shum, and
  Nikolay~I Zheludev.
\newblock An optical fiber network oracle for np-complete problems.
\newblock {\em Light: Science \& Applications}, 3(2):e147, 2014.

\bibitem{QI_bible}
Michael~A. Nielsen and Isaac~L. Chuang.
\newblock {\em Quantum Computation and Quantum Information}.
\newblock Cambridge Series on information and the Natural Sciences. Cambridge
  University Press, {10th Aniversary} edition, 2010.

\bibitem{Karp_10}
Richard~M Karp.
\newblock {\em 50 Years of Integer Programming 1958-2008}, chapter Reducibility
  among combinatorial problems, pages 219--241.
\newblock Springer Berlin/Heidelberg, 2010.

\bibitem{Shor_1}
P.~W. Shor.
\newblock Polynomial-time algorithms for prime factorization and discrete
  logarithms on a quantum computer.
\newblock {\em SIAM Journal on Computing}, 26(5):1484--1509, 1997.

\bibitem{NP_optical2}
Damien Woods and Thomas~J Naughton.
\newblock Optical computing: photonic neural networks.
\newblock {\em Nature Physics}, 8(4):257--259, 2012.

\bibitem{NP_quantum}
Tien~D Kieu.
\newblock Quantum algorithm for $\mathrm{H}$ilbert's tenth problem.
\newblock {\em International Journal of Theoretical Physics}, 42(7):1461--1478,
  2003.

\bibitem{NP_molecular}
Leonard~M Adleman.
\newblock Molecular computation of solutions to combinatorial problems.
\newblock {\em Science}, 266(5187):1021--1024, 1994.

\bibitem{NP_DNA}
Z~Ezziane.
\newblock Dna computing: applications and challenges.
\newblock {\em Nanotechnology}, 17(2):R27, 2006.

\bibitem{NP_light}
Mihai Oltean.
\newblock Solving the hamiltonian path problem with a light-based computer.
\newblock {\em Natural Computing}, 7(1):57--70, 2008.

\bibitem{algorithms_book}
Sanjoy Dasgupta, Christos Papadimitriou, and Umesh Vazirani.
\newblock {\em Algorithms}.
\newblock McGraw-Hill, New York, NY, 2008.

\bibitem{wiley_enc}
F.~Bonani, F.~Cappelluti, S.~D Guerrieri, and F.~L. Traversa.
\newblock {\em Wiley Encyclopedia of Electrical and Electronics Engineering},
  chapter Harmonic Balance Simulation and Analysis.
\newblock John Wiley and Sons, Inc., 2014.

\bibitem{Goertzel}
Gerald Goertzel.
\newblock An algorithm for the evaluation of finite trigonometric series.
\newblock {\em The American Mathematical Monthly}, 65(1):34--35, 1958.

\bibitem{Shannon}
C.~E. Shannon.
\newblock A mathematical theory of communication.
\newblock {\em The Bell System Technical Journal,}, 27:379--423, 623--656,
  1948.

\bibitem{Communication_book}
Herbert Taub and Donald~L. Schilling.
\newblock {\em Principles of Communication Systems}.
\newblock McGraw-Hill Higher Education, 2nd edition, 1986.

\bibitem{DCRAM}
Fabio~L. Traversa, Fabrizio Bonani, Yuriy~V. Pershin, and Massimiliano
  Di~Ventra.
\newblock Dynamic $\mathrm{C}$omputing $\mathrm{R}$andom $\mathrm{A}$ccess
  $\mathrm{M}$emory.
\newblock {\em Nanotechnology}, 25:285201, 2014.

\bibitem{07_waser}
Rainer Waser and Masakazu Aono.
\newblock Nanoionics-based resistive switching memories.
\newblock {\em Nature materials}, 6:833, 2007.

\bibitem{08_strukov}
Dmitri~B. Strukov, Gregory~S. Snider, Duncan~R. Stewart, and R.~Stanley
  Williams.
\newblock {The missing memristor found}.
\newblock {\em Nature}, 453(7191):80--83, May 2008.

\bibitem{09_memory_materials}
T.~Driscoll, Hyun-Tak Kim, Byung-Gyu Chae, Bong-Jun Kim, Yong-Wook Lee,
  N.~Marie Jokerst, S.~Palit, D.~R. Smith, M.~Di~Ventra, and D.~N. Basov.
\newblock Memory metamaterials.
\newblock {\em Science}, 325(5947):1518--1521, 2009.

\bibitem{12_grollier}
A.~Chanthbouala, V~Garcia, R.~O. Cherifi, K.~Bouzehouane, S~Fusil, X.~Moya,
  S.~Xavier, H.~Yamada, C.~Deranlot, N.~D. Mathur, M.~Bibes, A~Barth�l�my,
  and J.~Grollier.
\newblock A ferroelectric memristor.
\newblock {\em Nature materials}, 11:860--864, 2012.

\end{thebibliography}


\section*{\large  Supplementary Information}

\begin{figure*}
\begin{center}
\includegraphics[width=1.6\columnwidth]{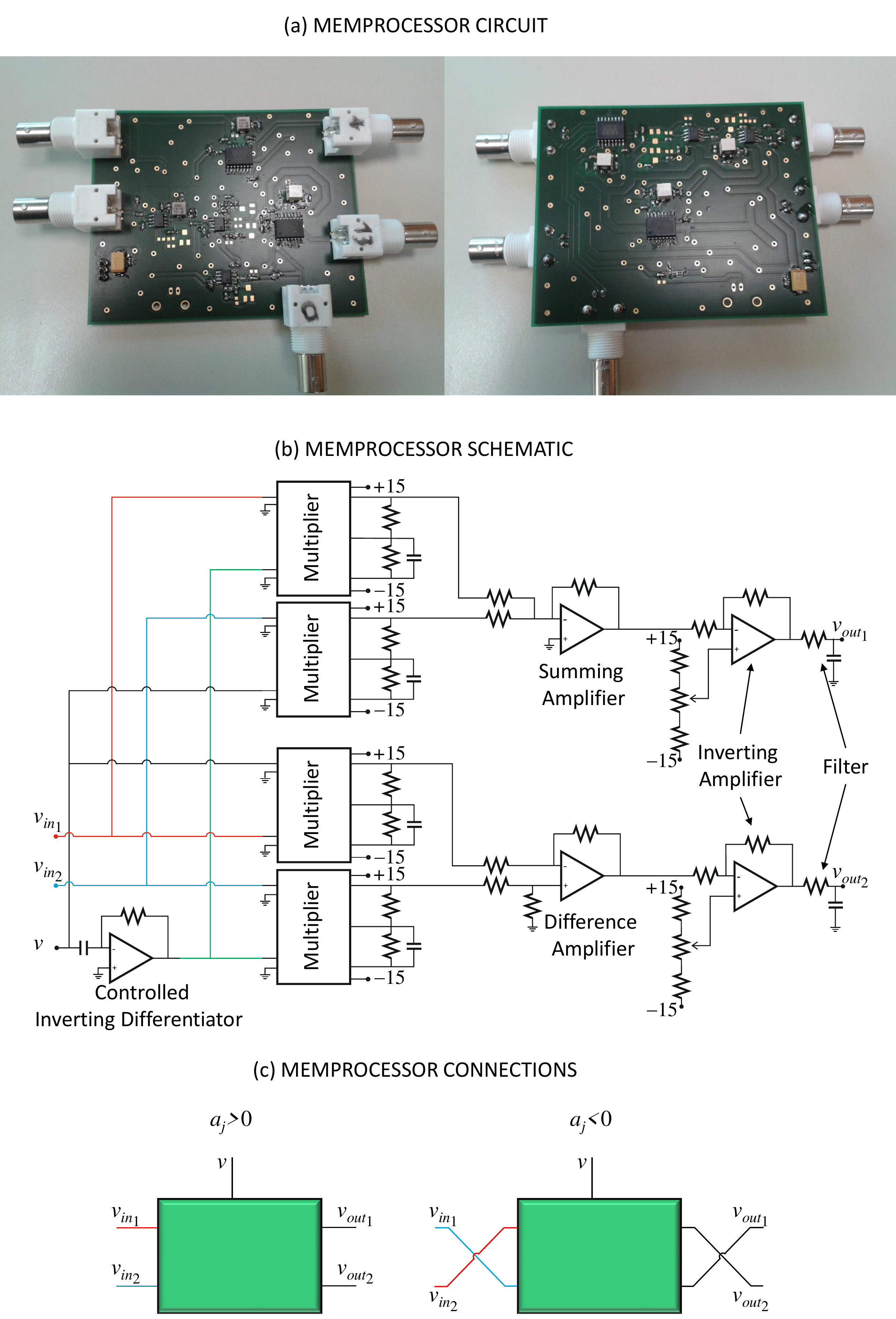}
\end{center}
\caption{\label{memprocessor_SI}Memprocessor module. (a) Pictures, (b) schematic, (c) connection rule of the memprocessor module.}
\end{figure*}

\begin{figure*}
\begin{center}
\includegraphics[width=1.6\columnwidth]{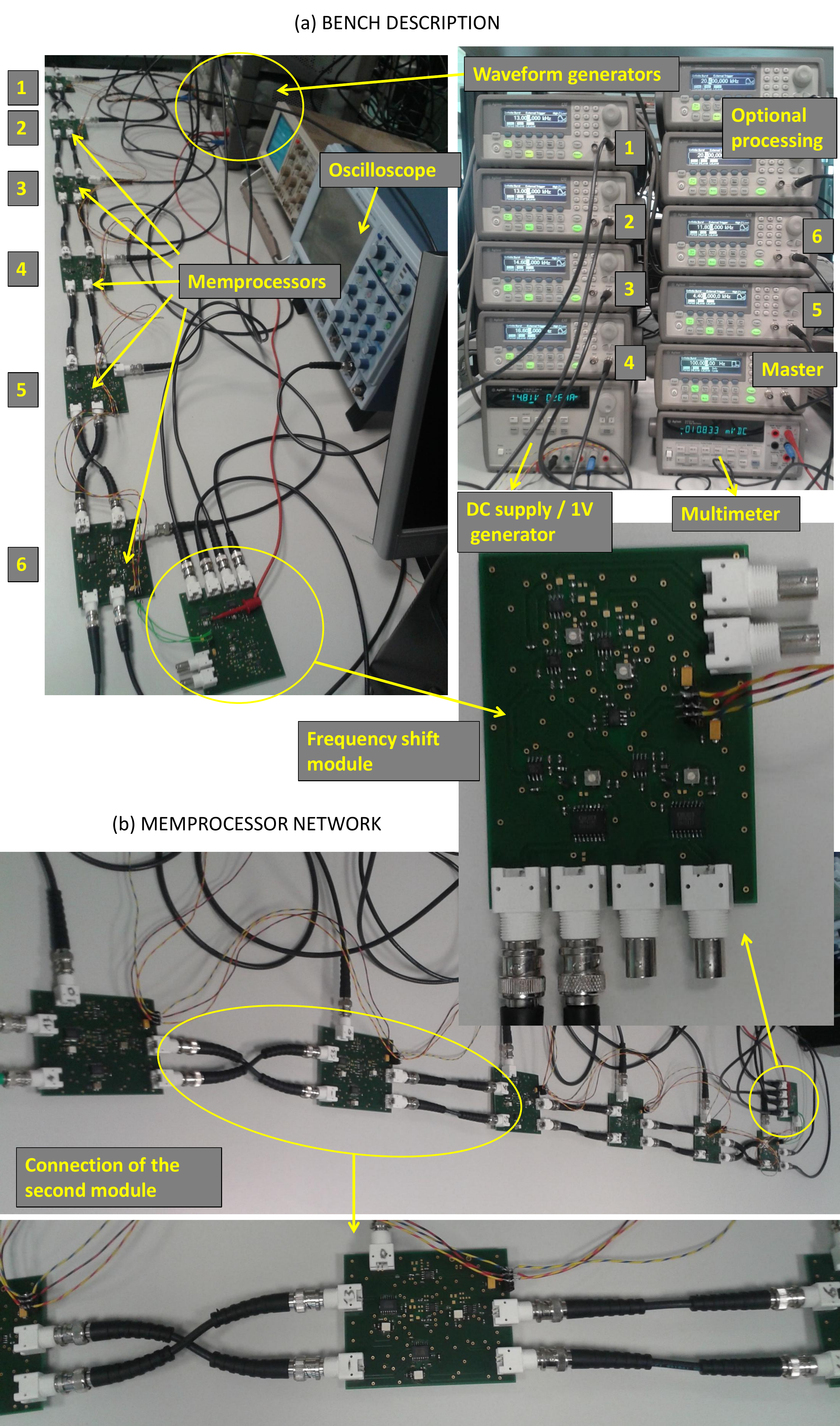}
\end{center}
\caption{\label{schemabanco}(a) Sketch of the laboratory test bench. (b) Synchronization of the waveform generators. (c) Collective state of 6 memprocessors network read by the oscilloscope.}
\end{figure*}

\begin{figure*}
\begin{center}
\includegraphics[width=1.6\columnwidth]{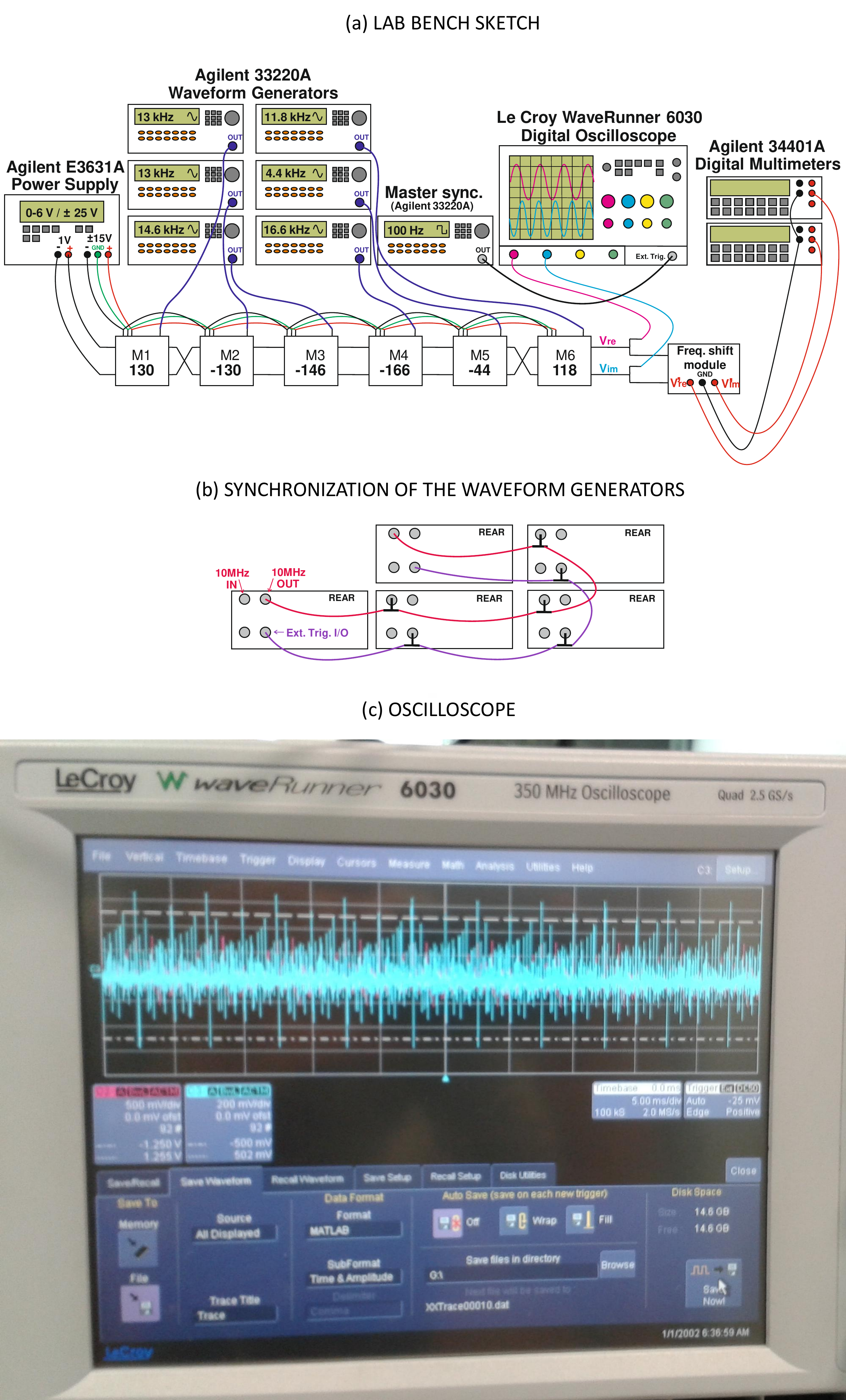}
\end{center}
\caption{\label{fotobanco}Pictures of the laboratory test bench.}
\end{figure*}

\end{document}